\documentclass[twocolumn,showpacs,showkeys,superscriptaddress,aps,pra]{revtex4}

\usepackage{amsmath}
\usepackage{amssymb}
\usepackage{color}
\usepackage{verbatim}
\usepackage{array}
\usepackage{graphicx}
\usepackage{epsfig}
\usepackage{bm}
\usepackage[ragged]{sidecap}

\newcommand{\ds}{\displaystyle}
\newcommand{\ddsum}[1]{{\displaystyle \sum_{ #1 }}}

\newcommand{\supercomas}[1]{``#1''}
\newcommand{\Red}[1]{{\color{black}{#1}}}
\newcommand{\Blue}[1]{{\color{black}{ #1}}}
\def\bra#1{\mathinner{\langle{#1}|}}
\def\ket#1{\mathinner{|{#1}\rangle}}
\newcommand{\braket}[2]{\langle #1|#2\rangle}

\newcommand{\dd}{\mathrm{d}}
\newcommand{\ti}{~}

\renewcommand{\vec}{\bm}
\begin{document}
\preprint{\hfill\parbox[b]{0.3\hsize}{ }}

%=========Title=======================================================
\title{Angular and polarization analysis for two-photon decay of $2s$ hyperfine states of \Blue{hydrogenlike Uranium}}

%=========Authors=====================================================
\author{L. Safari}
\thanks{laleh.safari@oulu.fi
}
\affiliation{Department of Physics, University of Oulu, Fin-90014 Oulu, Finland}

\author{P. Amaro}
%\affiliation{Physikalisches Institut, Ruprecht-Karls-Universit\"{a}t Heidelberg, D-69120 Heidelberg, Germany}
\affiliation{Centro de F\'{i}sica At\'{o}mica, Departamento de F\'{i}sica, Faculdade de Ci\^{e}ncias e Tecnologia, FCT, Universidade Nova de Lisboa, P-2829-516 Caparica, Portugal}

\author{J. P. Santos}
\affiliation{Centro de F\'{i}sica At\'{o}mica, Departamento de F\'{i}sica, Faculdade de Ci\^{e}ncias e Tecnologia, FCT, Universidade Nova de Lisboa, P-2829-516 Caparica, Portugal}

\author{F. Fratini}
%\email[]{ffratini@fisica.ufmg.br}
\affiliation{Department of Physics, University of Oulu, Fin-90014 Oulu, Finland}
\affiliation{Departamento de F\'isica, Universidade Federal de Minas Gerais, 31270-901 Belo Horizonte, MG, Brasil}
\affiliation{Institut N\'eel-CNRS, BP 166, 25 rue des Martyrs, 38042 Grenoble Cedex 9, France}

\date{\today \\[0.3cm]}%
%=========Abstract====================================================
\begin{abstract}
\Red{The amplitude of two-photon transitions between hyperfine states in hydrogenlike ions is derived based on relativistic Dirac equation and second order perturbation theory. We study angular and linear polarization properties of the photon pair emitted in the decay of $2s$ states, where spin-flip and non-spin-flip transitions are highlighted. We pay particular attention to hydrogenlike uranium, since it is an ideal candidate for investigating relativistic and high-multipole effects, such as spin-flip transitions}. Two types of emission patterns are identified: i) non-spin-flip transitions are found to be characterized by an angular distribution of the type $W(\theta)\sim1+\cos^2\theta$ while the polarizations of the emitted photons are parallel; ii) spin-flip transitions have somewhat smaller decay rates and are found to be characterized by an angular distribution of the type $W(\theta)\sim1-1/3\cos^2\theta$ while the polarizations of the emitted photons are orthogonal, where $\theta$ is the angle between photons directions. Deviations due to non-dipole and relativistic contributions are evaluated for both types of transitions. This work is the first step toward exploring the effect of nucleus over the the angular and polarization properties of the photon pairs emitted by two-photon transitions.

\end{abstract}

%=========Classification==============================================

%=========Text========================================================
\maketitle

%
%
%
%
% ------------------------------ Introduction -------------------------- %

\section {Introduction} 

Two-photon decay in atoms and ions was introduced by  Max Born's PhD student Goeppert-Mayer in 1931 \cite{GoM31}.
Since then, many aspects of such a process, like the total decay rate and the spectral distribution, have been extensively investigated in the context of few-electron atoms and ions, both in theory and experiments \cite{Sh58, Dra86, Go81, DeJ97, SaP98, DuB93, AlA97, ScM99, MoD04, KuT09, Se10}.
Recently, some interest has been also devoted to the relativistic effects on angular and polarization properties of the two emitted photons \cite{Au76, An05, Les06, An09, An10, PolPol, Ped2012}, and to electron-electron interaction effects on the total decay rate \cite{Vol2011}. Apart from fundamental interest, two-photon transitions revealed themselves as a useful tool for investigation of different physical areas and applied science.
Already in 1940, for instance, Breit and Teller derived that the double photon emission was the principal cause of the decay of interstellar hydrogen atoms from their metastable $2s$-state \cite{Br40}, while, more recently, polarization properties of the emitted photons have been employed to successfully explore quantum entanglement  \cite{PiK85,KlD97,QuantInf}. 
Furthermore, two-photon transitions have been proposed as a tool to measure weak interaction properties \cite{Sch89, PNC}. 

In this article, the angular and polarization properties of the photon pair emitted by the two-photon decay of $2s$ hyperfine states in hydrogenlike ions are presented. \Red{Particular attention is paid to hydrogenlike uranium (U$^{91+}$). In fact, due to its strong electromagnetic field, high multipoles contributions that lead to spin-flip transitions are enhanced in this system. This makes hydrogenlike Uranium an ideal candidate for our studies.} The angular and polarization analysis of the emitted light is carried out within the Independent Particle Approximation (IPA), i.e., by coupling the spin-angular momenta of electron and nucleus and by neglecting any hyperfine interaction between electron and nucleus. \Red{We derive the analytical expression for the transitions amplitude within IPA.}
This work is a first step toward exploring the effect of the nuclear angular momentum (spin) on the angular and polarization properties of the emitted photons. \Red{Our analysis may pave the way for a new route to get information on the direction and the magnitude of the spin distribution inside the nucleus (which is still quite an unraveled problem) by using two-photon angular and polarization correlations.}

\section{Theoretical background}

% ------------------------------ Theory -------------------------- %
%
%
\subsection{Construction of the overall set of states} 

The presence of the nuclear spin has a twofold effect on the states of hydrogenlike systems. First, the energies of the atomic metastable states are slightly shifted, mainly due to the magnetic dipole interaction that nucleus and electron experience. This energy correction can be described by using first order perturbation theory with additional contributions, such as the relativistic, Bohr-Weisskopf, Breit-Rosenthal and QED contributions \cite{Sh97, Sa08}. Since this energy correction does not influence the angular and polarization properties of the emitted radiation, it will be totally neglected in the following. Second, the atomic states acquire a new quantum number, usually denoted by $F$, that represents the total angular momentum of the overall --nucleus plus electron-- system.

The overall atomic state can be described by coupling the nucleus and electron angular momenta (referred to as IPA), i.e by
\begin{equation}
\begin{array}{l}
\ket{n, \beta; F, I, \kappa, m_F}=\\[0.4cm]
\qquad \ddsum{m_I,\,m_j} \braket{j,m_j,I,m_I}{F,m_F} \ket{n; \kappa, m_j}\ket{\beta; I, m_I} ~,
\end{array}
\label{eq:totstate1}
\end{equation}
where $n$, $\kappa$ and $j$ are the (Bohr) principal, the Dirac and the angular momentum quantum numbers of the electron respectively, while $I$ represents the nuclear spin. On the other hand, $m_I$, $m_j$ and $m_F$ are the projections of the nuclear, electronic and total (nucleus plus electron) angular momenta onto the quantization axis, respectively.
Finally, $\beta$ is a collective label that denotes any other quantum number needed to specify the nuclear state
apart from $I$ and $m_{I}$.
Using standard notation, $\braket{j,m_j,I,m_I}{F,m_F}$ are Clebsch-Gordan coefficients.

To further proceed, we suppose that the nucleus does not interact with the radiation field. In the language of quantum mechanics, this equates to considering that the interaction Hamiltonian couples only electron fields through the photon emission, while it does not act on the quantum space of nuclear states.
This hypothesis holds for decays which involve bound states of neutral atoms, since the energy released in such decays is far lower than the nuclear excitation energies (there are few exceptions to this, like the nucleus of Th$^{229}$, where the first metastable excited state is $\lesssim 10$ eV above the ground state). For highly charged ions, nuclear excitations are of the order of $\sim$ MeV while photon energies can take values up to a hundred of keV. 
%------IMP
This supposition represents, therefore, a first approximation, since the inclusion of nuclear-field interactions in atomic decays is anyhow of higher order in perturbation theory \cite{Pa210}. 
%This is justified since the angular and polarization properties of the two emitted photons are mostly affected by the coupling of the angular momenta. In fact, as 
%------IMP
As a result of this assumption, we shall find in the next subsection that the radial part of the decay amplitude is characterized by only electron state components. On the other hand, we shall see that the angular part of the decay amplitude is characterized by both electron and nucleus states components, due to the coupling of their angular momenta. 
\Red{We shall see that the value for total spin quantum number will directly determine the shape of the angular and polarization distributions in the atomic transitions}.

\subsection{Second order transition amplitude} 

The theory of two-photon decay is based on the second-order transition amplitude and has been discussed in a number of recent papers \cite{PolPol, PNC, An10}. One of the characteristic features of such amplitude is that it contains a summation over the intermediate atomic states which runs through the whole atomic spectrum, including a summation over the discrete part as well as an integration over the (positive and negative) continuum. For the problem under consideration, such a summation splits up into summations over: i) the principal quantum number $n_{\nu}$, ii) the Dirac quantum number $\kappa_{\nu}$, iii) the total angular momentum $F_\nu$, and iv) its projection onto the quantization axis $m_{F_{\nu}}$. 

By using (\ref{eq:totstate1}) and by taking into account the orthonormality of the nuclear states, the amplitude for two-photon transitions between hyperfine states takes the form
\begin{widetext}
\begin{equation}
\begin{array}{c}
\ds\mathcal{M}^{\lambda_1\lambda_2}(i\to f) =
-(2\pi)\ddsum{T\,T'}\;\ddsum{\substack{\kappa_{\nu}\\m_I\,m_{j_{\nu}}}}\;
\ddsum{\substack{L_1\,L_2\\M_1\,M_2}}\;\ddsum{p_1\,p_2}\;\ddsum{\Lambda_1\,\Lambda_2}
(\lambda_1)^{p_1}(\lambda_2)^{p_2}[L_1,L_2]^{1/2}
i^{-L_1-L_2-p_1-p_2}\, \xi_{L_1\,\Lambda_1}^{p_1}
\xi_{L_2\,\Lambda_2}^{p_2} P^T \; P^{T'} \\[0.4cm]
\times\;\ds D^{L_2\,*}_{M_2\,\lambda_2}(\varphi_2,\theta_2,0)D^{L_1\,*}_{M_1\,\lambda_1}(\varphi_1,\theta_1,0) 
\Bigg[ 
U^{TT'}_{\Lambda_1\,\Lambda_2} \; \chi_{m_I\,m_{j_{\nu}}}^{f^T\, \nu^T} \chi_{m_I\,m_{j_{\nu}}}^{\nu^{T'}\, i^{T'}} 
\;+\; \big(1\leftrightarrow 2\big) \Bigg] ~,
\end{array} 
\label{Mfi2}
\end{equation}
\end{widetext}
where $\lambda_j$ and $\vec k_{j}$ are the helicity and wavevector of the $j$th photon. The term $D_{M_j\,\lambda_{j}}^{L_j}(\theta_j, \varphi_j)$ stands for the Wigner rotation matrices of order $L$ with angle cordinates $(\theta_j, \varphi_j)$. The notation $[L]$ stands for $2L+1$ and $\Lambda_{j}$ runs from $L_{j}-1$ to $L_{j}+1$. $T,T'=L,S$ denote the large (L) and small (S) components of the electron Dirac spinor, for which the factor $P^T$ is defined as $P^{L}=1$ and $P^{S}=-1$. Furthermore, $p_{1,2}=0,1$ and the function $\xi_{L\,\Lambda}^p$ is given by
\begin{equation}
\begin{array}{l c l}
\xi_{L\,\Lambda}^0&=&\delta_{L,\,\Lambda} \qquad,\\[0.2cm]
\xi_{L\,\Lambda}^1&=&\left\{
\begin{array}{l r}
\sqrt{\frac{L+1}{2L+1}} & \textrm{for } \Lambda=L-1\\[0.2cm]
-\sqrt{\frac{L}{2L+1}} & \textrm{for } \Lambda=L+1\\[0.2cm]
0 & \textrm{otherwise } ~.
\end{array}
\right.
\end{array}  
\label{eq:csi}
\end{equation}
The radial part of the amplitude in Eq.~(\ref{Mfi2}) is represented by the integral $U^{TT'}_{\Lambda_1\,\Lambda_2}$, which reads
\begin{equation}
U^{TT'}_{\Lambda_1\,\Lambda_2}=
\int \dd r \dd r' r^2r'^2 j_{\Lambda_1}(k_1r')j_{\Lambda_2}(k_2r) 
g_f^{\bar{T}*} g_{E_i+\omega_1}^{T\,\bar{T'}} g_i^{T'} ~,
\label{eq:int}
\end{equation}
where $g_{f,i}^T$ are the small and large radial components of the final and initial electron state, while
\begin{equation}
g_{E_i+\omega_1}^{T\,\bar{T'}}=\ddsum{n_{\nu}}\frac{g_{\nu}^T\,g_{\nu}^{\bar{T'}*}}{E_{\nu}-E_i-\omega_1}~,
\label{eq:green}
\end{equation}
is the radial Green function of the process. Here $E_{i,\nu}$ are the energies of the initial and intermediate atomic states, while $\bar T$ refers to the reverse radial component of $T$, $i.e.$ $\bar T=L$ for $T=S$ and vice versa.

The integral in Eq.~(\ref{eq:int}) involves only electron state components. However, its evaluation is not an easy task due to the (infinite) summation over the principal quantum number $n_{\nu}$ contained in the radial Green function. In the present work, such integral has been computed by using the Greens library \cite{Ko03}. Other computational techniques could be also used, such as B spline finite basis set method \cite{L.Safari:2012A,L.Safari:2012B}
The angular part of the amplitude in Eq.\ti(\ref{Mfi2}) is represented by the elements $\chi_{m_I\,m_{j_{\nu}}}^{f^T\, \nu^T}$ and $\chi_{m_I\,m_{j_{\nu}}}^{\nu^{T'}\, i^{T'}}$ therein contained and can be computed analytically:
\begin{equation}
\begin{array}{l}
\chi_{m_I\,m_{j_{\nu}}}^{f^T\, \nu^T} = \ddsum{m_{j_f}}\braket{j_f,m_{j_f},I,m_I}{F_f,m_{F_f}} \\
\qquad\times\;
\bra{\kappa_f, l_f^T, m_{j_f}} \vec\sigma\cdot\vec T^*_{L_2\,\Lambda_2\,M_2}\ket{\kappa_{\nu}, l_{\nu}^{\bar T}, m_{j_{\nu}}}\\[0.4cm]
\chi_{m_I\,m_{j_{\nu}}}^{\nu^{T'}\, i^{T'}} = \ddsum{m_{j_i}}\braket{j_i, m_{j_i}, I, m_I}{F_i, m_{F_i}} \\
\qquad \times\;
\bra{\kappa_{\nu}, l_{\nu}^{T'}, m_{j_{\nu}}}\vec\sigma\cdot\vec T_{L_1\,\Lambda_1\,M_1}^*\ket{\kappa_i, l_i^{\bar T'}, m_{j_i}} ~,
\end{array}
\label{eq:angpart}
\end{equation}
where $\vec\sigma$ are Pauli matrices. The elements $\bra{\kappa_f,l_f^{T},m_{j_f}}$
$\vec\sigma\cdot\vec T_{L_2\,\Lambda_2\,M_2}^* $
$\ket{\kappa_{\nu},l_{\nu}^{\bar{T}},m_{j_{\nu}}}$
and $\bra{\kappa_{\nu}, l_{\nu}^{T'}, m_{j_{\nu}}}$
$\vec\sigma\cdot\vec T_{L_1\,\Lambda_1\,M_1}^*$
$\ket{\kappa_i, l_i^{\bar T'}, m_{j_i}}$
have been already discussed elsewhere \cite{An05, An02} and will not be here recalled. 

\begin{figure}[t]
\centering
\includegraphics[width=.48\textwidth]{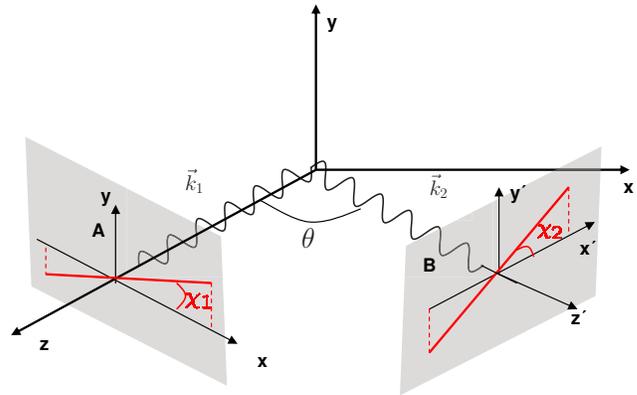}
\caption{(color online). Geometry considered for the two-photon emission. The propagation direction of the first
photon is adopted as $z$ direction. $x$ is chosen such that $xz$ is the reaction plane 
(plane spanned by the photons directions). $\theta$ is the angle between the photons directions, while angles $\chi_{1,2}$
define the linear polarizations of the first and second photon respectively with respect to their respective polarization planes. 
The polarization plane of the first (second) photon is denoted by $A$ ($B$) and represents the plane 
orthogonal to the photon direction.
}
\label{fig:fig1}
\end{figure}

The initial and final states involved in the two-photon transitions which we shall analyze below are $unpolarized$. It has been shown that, for this case, we may arbitrarily adopt the quantization axis ($\hat z$) along the momentum of the \supercomas{first} photon: $\hat z\parallel \hat{k}_1$ \cite{An05, Les06, An09, An10, PolPol}. We furthermore adopt $\hat x$ such that the $xz$-plane is the reaction plane (plane spanned by the photons directions). Figure \ref{fig:fig1} sketches the geometry we consider for the decay.
Within this geometry, the Wigner matrices in Eq.~(\ref{Mfi2}) simplify to 
$D^{L_1\,*}_{M_1\,\lambda_1}(\varphi_1, \theta_1, 0)=\delta_{M_1,\,\lambda_1}$ 
and
$D^{L_2\,*}_{M_2\,\lambda_2}(\varphi_2, \theta_2, 0)=d^{L_2}_{M_2\,\lambda_2}(\theta)$, 
where $d^{L}_{M\,\lambda}(\theta)$ is the reduced Wigner matrix and $\theta\equiv\theta_2$ is the polar angle of the second photon, which coincides, in the chosen geometry, with the angle between the photons directions (opening angle). Hence, the relative photons directions are uniquely determined by assigning the opening angle $\theta$, which will be the independent variable for plotting the angular distributions.

Since part of this work is devoted to analyze photons linear polarizations, further details concerning the detection geometry must be provided before proceeding with the analysis.
In Fig.~\ref{fig:fig1}, we show how the photon polarizations may be defined in a case experiment.
The polarization of each photon is measured in the \supercomas{polarization plane}, which is the plane orthogonal to the photon direction. In Fig.~\ref{fig:fig1}, the polarization planes of the first and second photon are denoted by $A$ and $B$, respectively.
Each detector is supposed to have a transmission axis, along which the linear polarization of the photon is measured. Such a transmission axis is rotated by an angle $\chi$ with respect to the reaction plane shown in Fig. \ref{fig:fig1} by the red dashed lines. Finally, each detector is supposed to work as a filter: Whenever a photon hits it, the detector either gives or does not give off a \supercomas{click}, which would respectively indicate that the photon has been measured as having its linear polarization along the direction $\chi$ or $\chi+90^{\circ}$. 

\subsection{Definition of angular and polarization correlations}

 Within IPA, equation (\ref{Mfi2}) represents the relativistic transition amplitude for the two-photon decay between hyperfine states in hydrogenlike ions. It contains the complete information on the emitted radiation.
Assuming that the ion is initially unpolarized and that the polarization of the final atomic state remains unobserved, taking into account the axes geometry chosen for the two-photon emission, and using the well-known relations between linear and circular polarization bases \cite{Ro53}, we can write the polarization-dependent differential decay rate as a function of the opening angle $\theta$ \cite{Go81}:
\begin{equation}
\begin{array}{l}
\ds W^{\chi_1\,\chi_2}(\theta)\equiv \frac{\dd w^{\chi_1 \chi_2}}{\dd \cos\theta }=
\frac{8\pi^2}{2F_i+1}
\ddsum{m_{F_i}\,m_{F_f}}\ddsum{\substack{\lambda_1\lambda_2\\\lambda_1'\lambda_2'}}\\[0.6cm]
\ds\qquad\times\int \dd\omega_1 \;\frac{\omega_1\omega_2}{4(2\pi)^3c^2}\; e^{i(\lambda_1-\lambda_1')\chi_1}
e^{i(\lambda_2-\lambda_2')\chi_2}\\[0.6cm]
\ds\qquad\times \;\mathcal{M}^{\lambda_1\lambda_2}
\mathcal{M}^{\lambda_1'\lambda_2'\,*}~.
\end{array}
\label{eq:ddrLP}
\end{equation}
In this article, the integration over the photon energies is numerically carried out by using the trapezoidal rule method. \Red{The number of points we used for the numerical integration has been checked to provide a precision of one percent}. Hereafter, the function $W^{\chi_1\,\chi_2}(\theta)$ shall be called \supercomas{polarization correlation}. It represents the probability density of detecting the emitted photons at the opening angle $\theta$ with defined linear polarizations $\chi_1$ and $\chi_2$.

Finally, by summing over the photons polarizations, we define the \supercomas{angular correlation} as
\begin{equation}
\begin{array}{l}
\ds W(\theta)\equiv\frac{\dd w}{\dd \cos\theta}=
\frac{8\pi^2}{2F_i+1}\int \dd \omega_1 \;
\frac{\omega_1\omega_2}{(2\pi)^3c^2}\\[0.4cm]
\qquad\times\;\ds\ddsum{m_{F_i}\,m_{F_f}}\ddsum{\lambda_1\lambda_2}
\Big|\mathcal{M}^{\lambda_1\lambda_2}(i\to f)\Big|^2 ~,
\end{array}
\label{eq:W}
\end{equation}
which represents the probability density of detecting the emitted photons at the opening angle $\theta$, irrespectively of their polarizations.

%
%
%
%
%
%
%
%
%
%
%
%
%
%
%
%
%
%
%
% ------------------------------ Results -------------------------- %
%
\begin{figure}
\centering
\includegraphics[width=.48\textwidth]{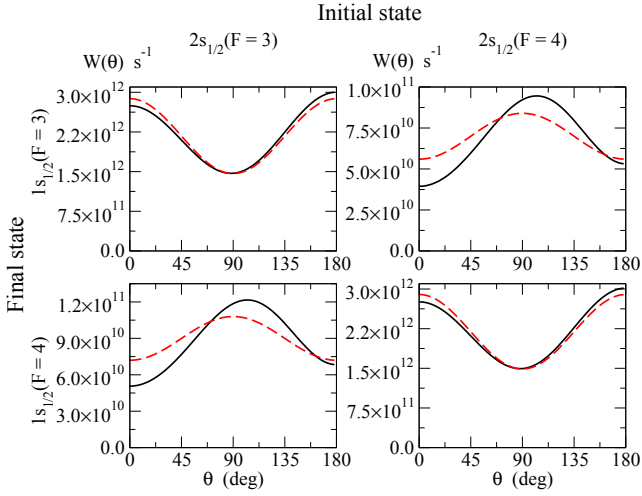}
\caption{(color online). Angular correlations in hydrogenlike ${}^{235}_{92}$U ions. The function $W(\theta)$ is shown for the transitions $2s_{1/2}\,(F=4,3)\to1s_{1/2}\,(F=4,3)$. The dashed-red curve refers to the electric dipole approximation while the solid-black curve refers to the full multipoles contribution. 
}
\label{fig:fig2}
\end{figure}
\section{Results and discussions}
\label{Res-Disc}

Here we analyze the angular and polarization correlations defined in Eqs.~\eqref{eq:W} and \eqref{eq:ddrLP} respectively, for decays of hyperfine $2s$ states in hydrogenlike ions, with special attention to hydrogenlike ${}^{235}_{92}$U ion, whose nuclear spin is I = 7/2 \cite{St05}. 
The function $W(\theta)$ obtained for $2s_{1/2}\,(F=4,3)\to1s_{1/2}\,(F=4,3)$ transitions in hydrogenlike ${}^{235}_{92}$U ion is displayed in Fig.~\ref{fig:fig2}. The full multipoles and the electric dipole (E1E1) contributions are separately displayed. 
Within the dipole approximation, the angular correlation for non-spin-flip transitions can be well described by the familiar shape $W_{E1E1}(\theta)\sim 1+\cos^2\theta$ \cite{MoD04}. However, the full-multipole calculation shows some asymmetric deviations from such a shape. This effect is already known from the past literature, where it has been showed that high multipoles contribute with terms of the type $\sim\cos\theta$ to the angular correlation in $2s_{1/2}\to1s_{1/2}$ transitions in highly charged ions \cite{Au76}. 
\begin{figure}[t]
\centering
\includegraphics[width=.48\textwidth]{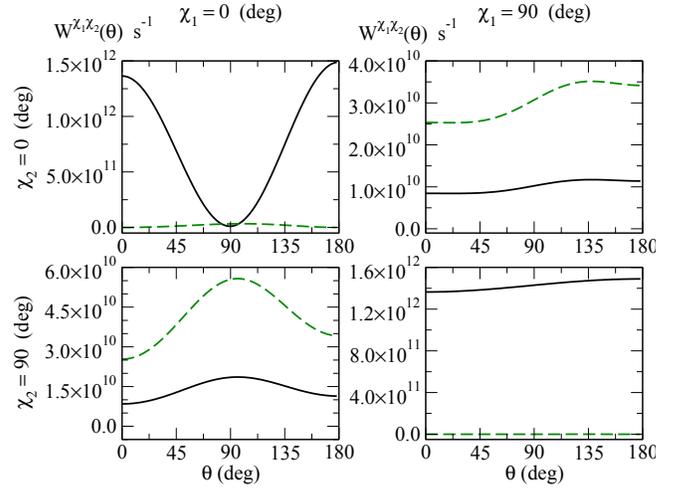}
\caption{(color online). Polarization correlations in hydrogenlike ${}^{235}_{92}$U ion. The function $W^{\chi_1\,\chi_2}(\theta)$ is shown for the transitions $2s_{1/2}\,(F=3)\to1s_{1/2}\,(F=3)$ (solid-black curve) and $2s_{1/2}\,(F=3)\to1s_{1/2}\,(F=4)$ (dashed-green curve). The four polarization configurations ($\chi_1, \chi_2=0^{\circ}, 90^{\circ}$) are displayed.
}
\label{fig:fig3}
\end{figure}

On the other hand, for spin-flip transitions, the angular correlation within the dipole approximation is well described by the function $W_{E1E1}(\theta)\sim 1-1/3\cos^2\theta$. This \Red{emission pattern} is typical for two-photon transitions of the type J$_{TOT}$ = 1 (0) $\to$ J$_{TOT}$ = 0 (1), where J$_{TOT}$ is the total angular momentum of the system which undergoes the decay. The two-photon decay $(1s\,2s)^3S_{J=1} \to (1s\,1s)^1S_{J=0}$ in heliumlike ions, where $J$ is the total angular momentum of the two-electron system, shows approximately the same behavior \cite{An10}.
As in the previous case, the full-multipole calculation shows remarkable asymmetric deviations from the symmetric shape. Quantitatively, the ratio $W(\pi)/W(0)$ is $\simeq 1.09$ for non-spin-flip transitions and $\simeq 1.35$ for spin-flip transitions.

For low-charged ions, we find that the angular correlation is fully described by the functions $\sim1+\cos^2\theta$ and $\sim1-1/3\cos^2\theta$ for non-spin-flip and spin-flip transitions, respectively (i.e., the full-multipole calculations coincides with the calculations performed within the dipole approximation).

From the figures, we also notice that spin-flip transitions are overall suppressed with respect to non-spin-flip transition, which is to be expected in view of the fact that the electric dipole, which is the leading multipole here, conserves the electron spin if evaluated nonrelativistically. This entails that the curves we obtained for spin-flip transitions are fully determined by relativistic and high-multipole contributions.

We now turn to analyze the polarization correlations for some of the hyperfine transitions considered in Fig. \ref{fig:fig2}. In Fig.~\ref{fig:fig3}, we plot the function $W^{\chi_1\chi_2}(\theta)$ as obtained for the transitions $2s_{1/2}(F=3)\to 1s_{1/2}(F=3)$ and $2s_{1/2}(F=3)\to 1s_{1/2}(F=4)$. We see that, in general, photons coming from spin-flip transitions and non-spin-flip transitions have mainly orthogonal and parallel linear polarizations respectively. We find that this polarization scheme holds perfectly (i.e., without deviations) for low-$Z$ hydrogenlike ions. However, for hydrogenlike Uranium, as well as for any highly charged ions, sizeable deviations are evident, as can be seen in Fig. \ref{fig:fig3}.

\Blue{Hydrogenlike heavy ions can be nowadays efficiently produced in storage rings \cite{A.Gumberidze:2013}}. The energy of the emitted radiation in hydrogenlike Uranium is in the range of hard X-rays. An experimental polarization analysis of such energetic photons would be nowadays possible through the use of Compton polarimeters 
\cite{TNS52a, TNS52b, RSI79, JI5, RSI67}.
\Blue{By analyzing the decay spectrum, a conventional photon-photon coincidence measurements enables one to distinguish two-photon decay events from the dominant single-photon M1 decay channel \cite{A.Simionovici:93,P.H.Mokler:04}.}
Therefore, 
%although Compton polarimeters are not normally used to record the polarization properties in multi-photon processes \cite{PRL97}, 
information on the polarization state of two photons can be 
%in principle 
achieved by selecting events which have been recorded in coincidence by two polarimeters 
and which have the desired scattering angle \cite{PRL97,PNC, NatWard, PRSny, PRWu, PRBleu}.

%\Blue{The $2s\to 1s$ transitions in heavy hydrogenlike ions is dominated by the single-photon M1 channel while the the two-photon channel characterizes only few percents of the transition events. Nevertheless, the contribution of the M1 channel can be clearly identified and subtracted in experiments by analyzing the spectrum of the transition \cite{A.Simionovici:93,P.H.Mokler:04}.}

\section{Summary and perspectives} 

In summary, the amplitude for two-photon transitions between hyperfine states in hydrogenlike atoms has been calculated. By using such amplitude, the angular and linear polarization properties of the photon pair emitted in two-photon decays of $2s$ hyperfine states have been investigated within second-order perturbation theory and the Dirac relativistic framework. Special attention has been paid to hydrogenlike ${}^{235}_{92}$U ion. Results have been showed for the transitions $2s_{1/2}\,(F=4,3)\to1s_{1/2}\,(F=4,3)$. It has been possible to identify two emission patterns: i) two-photon non-spin-flip transitions are found to be characterized by an angular distribution approximately of the type $W(\theta)\sim1+\cos^2\theta$ and by photon polarizations approximately parallel one to another; ii) two-photon spin-flip transitions have somewhat smaller decay rate and are found to be characterized by an angular distribution approximately of the type $W(\theta)\sim1-1/3\cos^2\theta$ as well as by photon polarizations approximately orthogonal one to another. Deviations to this patterns come from high-multipoles and relativistic contributions, are negligible for low-charged ions, and are of size 1 to 25\% in hydrogenlike ${}^{235}_{92}$U ion.

\Red{This article is the first step toward exploring the nuclear spin effect on the angular and polarization properties of the photon pair emitted in two-photon decays. This study might pave the way for a new route to get information on the direction and the magnitude of the spin distribution inside the nucleus, which is still quite an unraveled problem \cite{Jaf1995}}.

\section{Acknowledgments} 

F.F. acknowledges support by Funda\c{c}\~ao de Amparo \`a Pesquisa do estado de Minas Gerais (FAPEMIG) and Conselho Nacional de Desenvolvimento Cient\'ifico e Tecnol\'ogico (CNPq).
 F.F. and L.L. acknowledge support by the Research Council for Natural Sciences and Engineering of the Academy of Finland.
J. P. S. and P. A. acknowledge support by FCT -- Funda\c{c}\~ao para a Ci\^encia e a Tecnologia (Portugal), 
through the Projects No. PEstOE/FIS/UI0303/2011 and PTDC/FIS/117606/2010, financed by the European Community 
Fund FEDER through the COMPETE -- Competitiveness Factors Operational Programme.

%=========Bibliography========================================================

\end{document}